\documentclass[twocolumn,showpacs,preprintnumbers,amsmath,amssymb,superscriptaddress]{revtex4}

\usepackage{graphicx,dcolumn,bm}

\newcommand{\nl}{\nonumber \\}

\newcommand{\be}{\begin{equation}}
\newcommand{\ee}{\end{equation}}
\newcommand{\bea}{\begin{eqnarray}}
\newcommand{\eea}{\end{eqnarray}}
\newcommand{\bsube}{\begin{subequations}}
\newcommand{\esube}{\end{subequations}}
\newcommand{\Fig}[1]{Fig.\,\ref{#1}}
\newcommand{\Eq}[1]{Eq.\,(\ref{#1})}

\newcommand{\la}{\langle}
\newcommand{\ra}{\rangle}

\newcommand{\rmB}{{\rm B}}

\newcommand{\rmL}{{\rm L}}
\newcommand{\rmR}{{\rm R}}
\newcommand{\rmd}{{\rm d}}
\newcommand{\rmi}{{\rm i}}
\newcommand{\rmc}{{\rm c}}
\newcommand{\Tr}{{\rm Tr}}
\newcommand{\epl}{\epsilon}
\newcommand{\alf}{\alpha}
\newcommand{\sgm}{\sigma}
\newcommand{\upa}{\uparrow}
\newcommand{\omg}{\omega}
\newcommand{\Gam}{\Gamma}
\newcommand{\GamL}{\Gamma_{\rm L}}
\newcommand{\GamR}{\Gamma_{\rm R}}
\newcommand{\dwa}{\downarrow}
\newcommand{\vpl}{\varepsilon}
\newcommand{\Gup}{G(0)}

\begin{document}

\title{Conditional spin counting statistics as a probe of Coulomb interaction
and spin-resolved bunching}

\author{JunYan Luo}
\email{jyluo@zust.edu.cn}
\affiliation{School of Science, Zhejiang University of Science
  and Technology, Hangzhou 310023, China}
\author{Jing Hu}
\affiliation{School of Science, Zhejiang University of Science
  and Technology, Hangzhou 310023, China}
\author{Xiaoli Lang}
\affiliation{School of Science, Zhejiang University of Science
  and Technology, Hangzhou 310023, China}
\author{Yu Shen}
\affiliation{School of Science, Zhejiang University of Science
  and Technology, Hangzhou 310023, China}
\author{Xiao-Ling He}
\affiliation{School of Science, Zhejiang University of Science
  and Technology, Hangzhou 310023, China}
\author{HuJun Jiao}
\affiliation{Department of Physics, Shanxi University, Taiyuan,
 Shanxi 030006, China}

\date{\today}

 \begin{abstract}
 Full counting statistics is a powerful tool to characterize
 the noise and correlations in transport through mesoscopic
 systems.
 In this work, we propose the theory of conditional spin counting
 statistics, i.e., the statistical fluctuations of spin-up (down)
 current given the observation of the spin-down (up) current.
 In the context of transport through a single quantum dot,
 it is demonstrated that a strong Coulomb interaction
 leads to a conditional spin counting statistics that
 exhibits a substantial change in comparison to that
 without Coulomb repulsion.
 It thus can be served as an effective way to probe the
 Coulomb interactions in mesoscopic transport systems.
 In case of spin polarized transport, it is further shown that
 the conditional spin counting statistics offers a
 transparent tool to reveal the spin-resolved bunching
 behavior.
 \end{abstract}
 

 \maketitle



 \section{\label{thsec1}Introduction}

 The exploration of full counting statistics (FCS) in mesoscopic
 systems has vital roles to play in providing penetrating insight
 into microscopic mechanisms in transport and temporal
 correlations between charge carriers which are not
 accessible from the conventional measurements of
 time-averaged current alone \cite{Naz03,Bla0511478,Bla001}.
 Particularly, recent advances in nanotechnology
 have made it possible to measure electron transport processes
 that take place at single-electron level \cite{Lu03422,Fuj042343,Byl05361,%
 Elz04431,Sch042005,Fuj061634,Gus06076605,Suk07243,Gus09191,Cho12024273}.
 All statistical cumulants of the number of transferred
 particles can now be extracted experimentally.

 Theoretical study of FCS based on the scattering approach
 turns out to be very powerful for characterizing statistics
 of noninteracting  electron transport through various
 systems, such as normal-superconductor structures \cite{She03485,Lev04115305,Bel01067006,Bel01197006},
 chaotic cavities \cite{Pil03206801,Nag04176804,Pil04045304,%
 Nov07073304,Nov08035337}, and electron entanglement
 detection devices \cite{Sim06020407,Gio06115315,Gio07241305}.
 Yet,  with continued miniaturization of the
 system size, the involving many-particle interactions
 become increasingly important in mesoscopic
 transport \cite{Pee001023,Kou01701}.
 For that purpose, a generalized quantum master equation (QME) approach
 has been established by Bagrets and Nazarov, with the Coulomb
 interactions being fully taken into account \cite{Bag03085316}.
 This approach has been widely employed to analyze the FCS in a variety
 of structures, for instance, quantum dot (QD)
 systems \cite{Bel05161301,Kie06033312,Gro06125315,Wan07125416,%
 Wel08195315,Urb09165319,Xue13208},
 molecules \cite{Imu07205341,Xue10033707,Wey12205306,Xue11716,Xue11083706}, and
 nanoelectromechanical resonators \cite{Fli05411,Nov04248302,Rod05085324}.
 Furthermore, the QME approach was recently extended to investigate
 finite-frequency FCS \cite{Ema07161404,Mar11125426} as well as non-Markovian
 dynamics \cite{Agu04206601,Bra06026805,Fli08150601,Jin11053704,Luo13173703}.

 \begin{figure}
 \begin{center}
 \includegraphics*[scale=0.40]{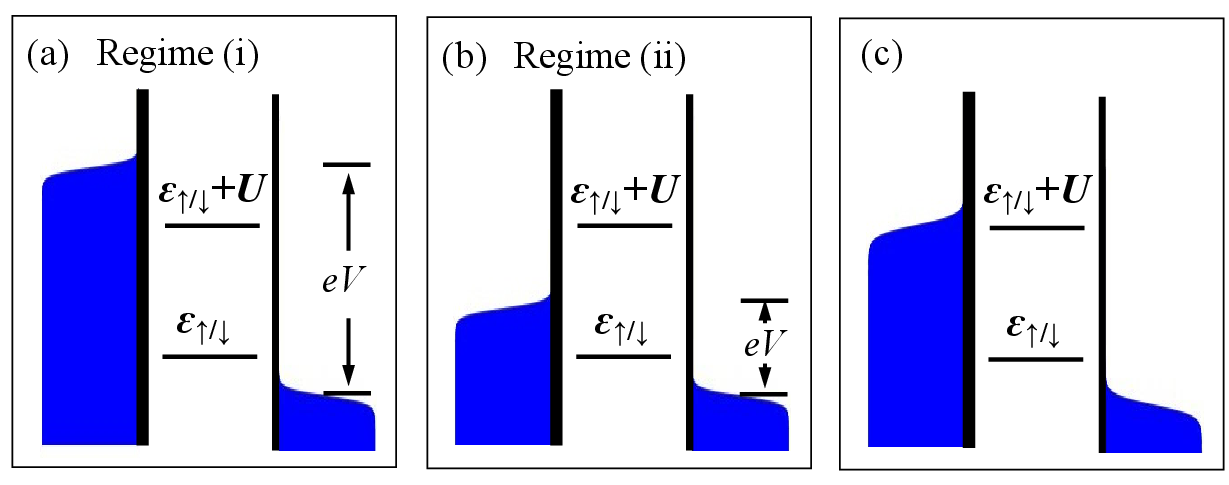}
 \caption{\label{Fig1}
 Schematic setup of transport through a single QD in
 different bias configurations.
 (a) Regime (i): The bias is large enough to overcome the
 Coulomb interaction such that the two excitation energies
 $\epl_\sgm$ and $\epl_\sgm+U$ are within the bias window.
 (b) Regime (ii): A small bias is applied across the QD, and
 only excitation energy $\epl_\sgm$ lies in the bias window.
 Double occupation on the QD is prohibited.
 (c) The bias voltage is applied in such a way that double occupation
 on the QD is partially allowed. It offers an interpolation
 between the regimes (i) and (ii).}
 \end{center}
 \end{figure}

 For statistically independent tunneling events, the
 current fluctuations exhibit Poissonian statistics.
 Normally, the presence of Pauli exclusion principle, which prohibits
 two fermions of the same spin to be superimposed, leads to the suppression
 of the current noise below the Poisson value \cite{Che914534,Mar921742}.
 On the other hand, Coulomb repulsion acts as another important
 correlation mechanism that might enhance or inhibit noise,
 depending on different physical regimes
 concerned \cite{Mar003386,Bla9910217,Suk01125315,Cot04206801,%
 Kie07206602,Urb08075318}.
 Yet, in reality it is quite difficult to distinguish the
 effects of Coulomb repulsion and the Pauli principle in the
 charge current noise.
 It is thus instructive to find a transparent and direct way
 to characterize the degree of Coulomb correlation in mesoscopic
 transport.
 For this purpose, we propose in this work the theory of conditional
 spin counting statistics:  The statistical fluctuations of spin up
 (down) current given the observation of the spin down (up) current.
 The inspiration of this theory comes from fact that the Pauli
 exclusion principle only acts on fermions of the same spin,
 while electrons with opposite spins are only correlated via the
 Coulomb repulsion.

 First, we consider electron transport through a single
 QD tunnel-coupled to two normal electrodes.
 Although the net current is spin unpolarized, the up
 and down spins are intrinsically correlated to each
 other via Coulomb repulsion.
 It is demonstrated that the Coulomb correlation gives
 rise to conditional spin counting statistics that
 exhibits a substantial change in comparison to the
 uncorrelated one.
 It thus may be utilized as an effective way to characterize
 the Coulomb correlation in various mesoscopic transport systems.

 Second, we investigate conditional spin counting statistics for
 spin polarized transport by taking into account ferromagnetic
 electrodes.
 It is worthwhile to mention that the (unconditional) spin
 counting statistics has been studied for many
 years \cite{Lor04046601,Sch07235105,Sau04106601,Luo08345215}.
 It was shown that spin current noise can be utilized to
 detect spin unit of quasiparticles \cite{Wan04153301},
 to sensitively probe spin decoherence in a spin
 battery \cite{Don05066601}, to reveal the discrete nature
 of the photon states for a quantum dot coupled to a
 cavity field \cite{Dju06115327}.
 In comparison with the unconditional spin counting statistics,
 we will show the conditional one may serve as a transparent and
 sensitive tool to investigate spin-resolved bunching behavior.


 The rest of the paper is organized as follows.
 In Section \ref{thsec2}, we describe the single QD system
 under different bias configurations, corresponding to different
 effectiveness of the Coulomb correlations.
 We discuss in Section \ref{thsec3} the charge FCS, which will be
 compared with the conditional spin counting statistics in
 sensing the Coulomb repulsion.
 Section \ref{thsec4} is devoted to the theory of conditional spin
 counting statistics.
 Its application to the single QD system
 is demonstrated in Section \ref{thsec5}, with focus on
 its effectiveness in characterizing Coulomb correlation and
 spin-resolved bunching behavior.
 It is then followed by the conclusion in Section \ref{thsec6}.

 \section{\label{thsec2}The Model}

 We consider electron transport through a single QD
 with Coulomb interaction, as schematically shown
 in \Fig{Fig1}. The entire system is described by the
 Hamiltonian $H=H_{\rmB}+H_{\rm QD}+H'$, with
 \bsube
 \bea
 &&H_{\rmB}=\sum_{\alpha=\rmL,\rmR}\sum_{k \sgm} \epl_{\alpha k\sgm}
 c_{\alpha k \sgm}^\dag c_{\alpha k \sgm},
 \\
 &&H_{\rm QD}=\sum_{\sgm}\vpl_{\sgm}d_\sgm^\dag d_\sgm+Un_\upa n_\dwa,
 \\
 &&H'=\sum_{\alpha=\rmL,\rmR}\sum_{k \sgm}(t_{\alpha k\sgm}
 c_{\alpha k \sgm}^\dag d_\sgm+{\rm h.c.}).
 \eea
 \esube
 Here $H_\rmB$ models the noninteracting electrons in the
 left ($\alpha$=L) and right ($\alpha$=R) electrodes,
 with $c_{\alpha k \sgm}^\dag$ ($ c_{\alpha k \sgm}$) the
 electron creation (annihilation) operator in the
 corresponding electrode.
 The electron distributions in the electrodes are governed
 by the electrochemical potentials $\mu_\rmL$ and $\mu_\rmR$,
 which define the voltage $eV=\mu_\rmL-\mu_\rmR$.
 $H_{\rm QD}$ describes the QD with one spin-degenerate
 energy level $\vpl_\sgm$ and the Coulomb interaction $U$
 on the dot, where $n_\sgm=d_\sgm^\dag d_\sgm$ is the
 occupation operator, with $d_\sgm^\dag$ ($d_\sgm$) the
 electron annihilation (creation) operator in the QD.
 Electron tunneling between electrodes and QD is depicted
 by $H'$. The tunneling rate for a spin-$\sgm$ electron is
 characterized by the intrinsic tunneling width
 $\Gam_{\alpha\sgm}(\omg)=2\pi\sum_k |t_{\alpha k\sgm}|^2\delta(\epl_{\alpha k\sgm}-\omg)$.
 Hereafter we consider normal electrodes, i.e.
 $\Gam_{\alpha\upa}=\Gam_{\alpha\dwa}$, and assume flat bands in the electrodes, which
 yields energy-independent couplings $\Gam_{\alpha\sgm}$.
 The total tunnel-coupling strength thus is given by
 $\Gam_{\alpha}=\Gam_{\alpha\upa}+\Gam_{\alpha\dwa}$.
 Throughout this work, we set $\hbar=1$ for the Planck
 constant, unless stated otherwise.

 \begin{figure}
 \begin{center}
 \includegraphics*[scale=0.8]{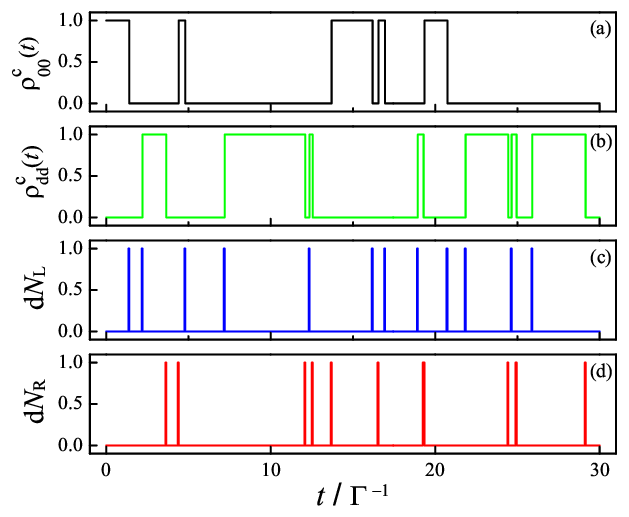}
 \caption{\label{fig2}
 Set of real-time density matrix elements $\rho^\rmc_{00}(t)$ and $\rho^\rmc_{\rm dd}(t)$
 for the regime (i), corresponding to the instantaneous states of empty and
 doubly occupied QD, respectively.
 The jumps in $\rho^\rmc_{00}(t)$ and $\rho^\rmc_{\rm dd}(t)$ are directly
 related to random electron process of tunneling into/out of the QD,
 represented by the stochastic variables d$N_{\rmL/\rmR}$.
 The unidirectional flow of electrons allows one to count electrons
 transporting through the system, leading thus to the spectrum of FCS.
 We assume symmetric tunnel-couplings, i.e. $\GamL=\GamR$, and
 $\Gam=\GamL+\GamR$.}
 \end{center}
 \end{figure}

 By specifying which excitation energies lie within the energy
 window defined by the Fermi levels $\mu_\rmL$ and $\mu_\rmR$,
 the following bias configurations will be considered.
 Regime (i): The bias is large enough to overcome the Coulomb
 interaction and thus the excitation energy levels $\epl_\sgm$
 and $\epl_\sgm+U$ are within the bias window defined by chemical potentials
 $\mu_\rmL$ and $\mu_\rmR$, as schematically shown in \Fig{Fig1}(a).
 The involving states include $|0\ra$-empty QD, $|\sgm\ra$-single
 occupation by a spin-$\sgm$ electron, and $|\rmd\ra$-double occupation.
 Regime (ii): Only the single level $\vpl_{\sgm}$
 is within the bias window, as shown in \Fig{Fig1}(b).
 The charge transport is maximally correlated.
 The states available are $|0\ra$-empty QD and
 $|\sgm\ra$-singly occupied by a spin-$\sgm$ electron.
 By appropriately applying the gate and bias voltages, the
 system can be tuned to the situation between the regimes (i)
 and (ii) as shown in \Fig{Fig1}(c).
 It allows us to analyze the effect of finite
 Coulomb correlation on the conditional spin counting
 statistics between the uncorrelated and maximally correlation
 cases.
 Our analysis is based on a second-order Born-Markov quantum
 master equation for $\Gam\ll k_{\rm B}T$, where the
 sequential tunneling processes play the dominant
 role \cite{Gur9615932,Gur08075325}.
 Higher order tunneling events, such as cotunneling, Kondo
 effect are thus suppressed.
 An approach of this type has been widely used in the literature
 for studying bias voltage dependent transport characteristics
 in various nanostructures \cite{Hau96}, such as
 single \cite{Bel05161301,Thi03115105,Kie03125320} or double
 QD \cite{Wun05205319},
 where typical step-like transport features were revealed.
 In spite of the simple model considered here, we will show that
 it is adequate to address the essence of the conditional
 spin counting statistics and its effectiveness of characterizing
 the Coulomb correlation and investigate spin-resolved bunching
 characteristics.

 \section{\label{thsec3}Charge Full Counting Statistics}

 In the single electron tunneling regime,
 an extra electron can inject into the QD from the left electrode,
 dwell in the QD for a certain amount of time before it escapes
 to the right electrode.
 This stochastic process produces intriguing signatures of the
 electronic conductor.
 To study the fluctuations involved in transport, we will utilize
 a Monte Carlo approach to simulate the individual
 electron tunneling events.
 We first introduce two stochastic variables
 d$N_{\rmL\sgm}(t)$ and d$N_{\rmR\sgm}(t)$
 (with values either 0 or 1) to represent, respectively,
 the numbers of spin-$\sgm$ electron
 injected into the QD from the left electrode and that escaped
 to the right electrode from the QD, during the small time
 interval d$t$.
 One then arrives at the following conditional QME \cite{Goa01125326}
 \bea
 \rmd \rho^\rmc&\!\!=\!\!&-\rmi {\cal L}\rho^\rmc(t)\rmd t
 -\sum_{\sgm=\upa,\dwa}\{\Gam_{\rmL\sgm}{\cal A}[d^\dag_{\sgm}]
 +\Gam_{\rmR\sgm}{\cal A}[d_{\sgm}]
 \nl
 &&-{\cal P}_{\rmL\sgm}(t)-{\cal P}_{\rmR\sgm}(t)\}\rho^\rmc(t)\rmd t
 \nl
 &&+\sum_{\sgm=\upa,\dwa}\rmd N_{\rmL\sgm}
 \left[\frac{\Gam_{\rmL\sgm}{\cal J}[d^\dag_{\sgm}]}{{\cal P}_{\rmL\sgm}(t)}-1\right]\rho^{\rmc}(t)
 \nl
 &&+\sum_{\sgm=\upa,\dwa}\rmd N_{\rmR\sgm}\left[\frac{\Gam_{\rmR\sgm}{\cal J}[d_{\sgm}]}{{\cal P}_{\rmR\sgm}(t)}-1\right]\rho^{\rmc}(t),\label{rhoc}
 \eea
 where the involving superoperators are defined as
 ${\cal L}\rho^\rmc\equiv[H_{\rm QD},\rho^\rmc]$,
 ${\cal J}[X]\rho^{\rmc}\equiv X\rho^{\rmc}X^\dag$ and
 ${\cal A}[X]\rho^{\rmc}\equiv\frac{1}{2}(X^\dag X\rho^{\rmc}+\rho^{\rmc}X^\dag X)$.
 The superscript ``c'' attached to the reduced density matrix
 denotes that the quantum state is conditioned on the measurement results.
 For single electron tunneling events (point process), the two
 classical random variables satisfy
 \bsube\label{dN}
 \bea
 E[\rmd N_{\rmL\sgm}(t)]={\cal P}_{\rmL\sgm}(t)\rmd t=\Tr\{ {\cal J}[\sqrt{\Gam_{\rmL\sgm}}d_{\sgm}^\dag]\rho^{\rmc}\}\rmd t,\label{dNL}
 \\
 E[\rmd N_{\rmR\sgm}(t)]={\cal P}_{\rmR\sgm}(t)\rmd t
 =\Tr\{ {\cal J}[\sqrt{\Gam_{\rmR\sgm}}d_{\sgm}]\rho^{\rmc}\}\rmd t,\label{dNR}
 \eea
 \esube
 where $E[X]$ stands for an ensemble average of a classical
 stochastic process $X$.
 \Eq{rhoc} implies that electron tunneling events
 condition the future evolution of the reduced density matrix,
 while \Eq{dN} indicates the instantaneous reduced density matrix
 conditions the detected electron tunneling events.
 Within this approach, one thus is capable of propagating the
 conditioned reduced density matrix ($\rho^\rmc$) and measurement
 record (d$N_{\alf\sgm}$) self-consistently.

 \begin{figure}
 \begin{center}
 \includegraphics*[scale=0.8]{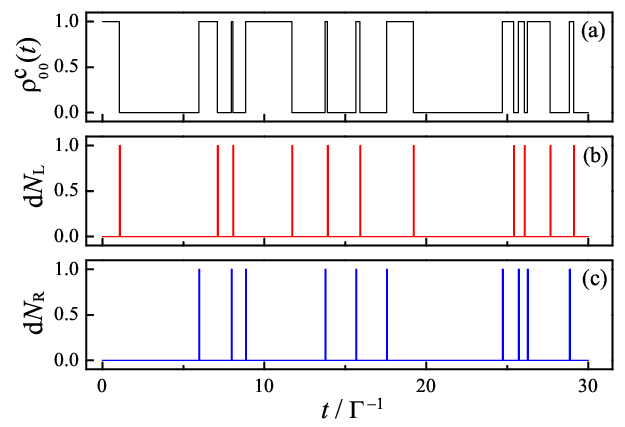}
 \caption{\label{fig3}
 Set of real-time quantum state $\rho^\rmc_{00}$ and corresponding
 detection records of tunneling into/out of the QD (d$N_{\rmL/\rmR}$)
 for the regime (ii), where double occupation on the QD is prohibited.
 The other parameters are the same as those in \Fig{fig2}.}
 \end{center}
 \end{figure}

 For the regime (i), the excitation energies $\epl_\sgm$ and
 $\epl_\sgm+U$ are within the bias window defined by the
 chemical potentials of the left and right electrodes.
 The QD can be empty, singly or doubly occupied.
 The instantaneous states of an empty [$\rho^{\rmc}_{00}(t)$]
 and doubly occupied [$\rho^{\rmc}_{\rm dd}(t)$] QD are
 displayed in \Fig{fig2}(a) and (b), respectively.
 A ``$1\rightarrow0$'' transition  in $\rho^{\rmc}_{00}$
 or the ``$0\rightarrow1$'' one in $\rho^{\rmc}_{\rm dd}(t)$
 indicates that an electron tunneled into QD from the left
 electrode (d$N_{\rmL}=1$), as shown in \Fig{fig2}(c).
 The opposite transitions ``$0\rightarrow1$'' in
 $\rho^{\rmc}_{00}$ and ``$1\rightarrow0$'' in
 $\rho^{\rmc}_{\rm dd}(t)$ imply tunneling events
 out of QD to the right electrode (d$N_{\rmR}=1$)
 [cf. \Fig{fig2}(d)].
 Here d$N_{\alpha}=\sum_{\sgm}$d$N_{\alpha\sgm}$ is the
 detected charge tunneling events, regardless of the spin
 orientations.
 The randomness in nonequilibrium charge transport are
 intimately related to the intriguing signatures of the
 electronic conductor, known as the full counting statistics.
 It can be described under the framework of counting
 field-dressed approach, in which electron tunnelings between
 reduced quantum system and the electrodes are taken into
 account by introducing a corresponding counting field $\chi$.
 This results in a $\chi$-resolved quantum master equation, which
 in the Born-Markov limit can be formally written
 as \cite{Lev964845,Agu04206601,Bra06026805,Fli08150601,Bag03085316}
 $\frac{\partial}{\partial t}{\rho}(\chi,t)={\cal L}_\chi \rho(\chi,t)$.
 All the cumulants of FCS can be obtained by taking derivatives
 of the cumulant generating function (CGF), which is determined
 from the lowest eigenvalue of ${\cal L}_\chi$. For instance, the CGF
 in the regime (i) is given by \cite{Bag03085316}
 \be\label{CGFI}
 {\cal F}_1(\chi)=-t_\rmc\{\GamL\!+\!\GamR
 -\sqrt{(\GamL\!-\!\GamR)^2+4\GamL\GamR e^{-\rmi \chi}}\},
 \ee
 where the counting time $t_\rmc$ satisfies
 $t_\rmc \gg \Gam_{\rmL/\rmR}^{-1}$.

 In the regime (ii), the excitation energy $\epl_\sgm+U$ stays well
 above the the Fermi level. Double occupation on the QD is prohibited,
 thus electrons can only transport through
 the single-level $\epl_\sgm$.
 The available Fock states
 are reduced to $|0\ra$-empty QD and $|\sgm\ra$-single occupation by a spin-$\sgm$
 electron.
 The real-time state $\rho_{00}^{\rmc}(t)$ is sufficient to characterize
 electron tunneling events as shown in \Fig{fig3}.
 Whenever $\rho_{00}^{\rmc}(t)$ jumps from 1 to 0, it reveals
 an electron has tunneled into the QD from the left electrode
 (d$N_{\rmL}=1$), while the `0$\rightarrow$1' jump implies
 the tunneling out event to the right electrode (d$N_{\rmR}=1$),
 as displayed in \Fig{fig3}(b) and (c), respectively.
 By counting the single electron tunneling events, one
 obtains the CGF in the regime (ii)
 \be\label{CGFII}
 {\cal F}_2(\chi)\!=\!-\frac{t_\rmc}{2}\{2\GamL\!+\!\GamR
 \!-\!\sqrt{(2\GamL\!-\!\GamR)^2\!+\!8\GamL\GamR e^{-\rmi \chi}}\}.
 \ee
 Apparently, the result for the regime (ii) [\Eq{CGFII}] is
 qualitatively similar to that of the regime (i) [\Eq{CGFI}],
 except for an effective doubling of the tunneling rate $\GamL$.
 This simple example shows that the charge FCS
 might not be the best possible probe of Coulomb interactions.
 On the contrary, we will show later that the intriguing behavior
 of conditional spin counting statistics can be utilized to reveal
 the Coulomb correlation between the up and down spins
 unambiguously.

 \section{\label{thsec4}Conditional Spin Counting Statistics}

 Now we introduce the theory of conditional spin counting
 statistics:
 The statistical fluctuations of the spin-$\upa$ ($\dwa$) current,
 given the observation of a given spin-$\dwa$ ($\upa$) current.
 It might be calculated from the conditional
 distribution function $P(I_\upa|I_\dwa)$ [$P(I_\dwa|I_\upa)$],
 the probability of observing spin-$\upa$ ($\dwa$) current conditioned
 on an observation of a spin-$\dwa$ ($\upa$) one.
 To obtain the mixed generating functions of conditional
 spin counting statistics, we introduce spin-resolved
 counting fields $\chi_\upa$ and $\chi_\dwa$, which are used
 to characterize, respectively, jumps of up and down spins
 through a specific junction.
 The corresponding ($\chi_\upa$,$\chi_\dwa$)-resolved
 master equation can be formally written as $\frac{\partial}{\partial t}
 \rho(\chi_\upa,\chi_\dwa,t)=M(\chi_\upa,\chi_\dwa)\rho(\chi_\upa,\chi_\dwa,t)$.
 In the steady state ($t_{\rm c}\gg\Gam_{\rmL/\rmR}^{-1}$),
 the joint generating function ${\cal F}(\chi_\upa,\chi_\dwa$)
 is determined from the
 minimal eigenvalue of $M$ according to
 \be
 {\cal F}(\chi_\upa,\chi_\dwa)=-\lambda_{\rm min}(\chi_\upa,\chi_\dwa) t_{\rm c},
 \ee
 where $\lambda_{\rm min}$ satisfies
 $\lambda_{\rm min}(\chi_\upa\rightarrow0,\chi_\dwa\rightarrow0)\rightarrow0$.

 From the definition of the joint generating function, the joint
 probability distribution of transmitted spins can be extracted
 by Fourier transforming on both variables,
 \be
 P(N_\upa,N_\dwa,t_{\rm c})=\!\int_0^{2\pi}\! \frac{\rmd \chi_\upa \rmd \chi_\dwa}{(2\pi)^2}
 e^{-\lambda_{\rm min}t_{\rm c}-\rmi(N_\upa \chi_\upa+N_\dwa \chi_\dwa)},
 \ee
 where $N_\upa=I_\upa t_{\rm c}$ and $N_\dwa=I_\dwa t_{\rm c}$.
 In the stationary limit, it is justified
 to evaluate the integral in the saddle point
 approximation \cite{Bag03085316,Pil03206801,Uts06086803,Uts07035333}. The
 dominant term contributing to the joint distribution is then given by
 \be\label{JPro}
 P(I_\upa,I_\dwa,t_{\rm c})=-t_{\rm c}{\rm min}_{\chi_\upa,\chi_\dwa}
 \{\lambda_{\rm min}+\rmi I_\upa \chi_\upa+\rmi I_\dwa \chi_\dwa\}.
 \ee
 The mixed generating functions of conditional spin counting statistics
 may be calculated by only Fourier transforming on one of the above
 variables in \Eq{JPro}.
 For instance, integrating over the $\chi_\upa$ variable yields
 ${\cal F}(\chi_\upa,I_\dwa)$.
 Due to the Bayes theorem, which relates the joint distribution
 and conditional distribution functions
 $P(I_\upa,I_\dwa)=P(I_\upa|I_\dwa)P(I_\dwa)=P(I_\dwa|I_\upa)P(I_\upa)$,
 the mixed generating function then is given by
 ${\cal F}(\chi_\upa|I_\dwa)={\cal F}(\chi_\upa,I_\dwa)-{\cal F}(0,I_\dwa)$.
 The statistical fluctuations of spin $\upa$ current given
 the observation of a spin $\dwa$ current can be obtained
 from ${\cal F}(\chi_\upa|I_\dwa)$ by performing derivatives
 with respect to the counting field $\chi_\upa$
 \be\label{CSCS}
 \la I_\upa^k\ra_{I_\dwa}=-\frac{(-\rmi\partial_{\chi_\upa})^k}{t_\rmc}
 {\cal F}(\chi_\upa|I_\dwa)|_{\chi_\upa\rightarrow0}.
 \ee
 Analogously, the conditional spin counting statistics of
 spin $\dwa$ current can be obtained from ${\cal F}(\chi_\dwa|I_\upa)$.

 \begin{figure}
 \begin{center}
 \includegraphics*[scale=0.8]{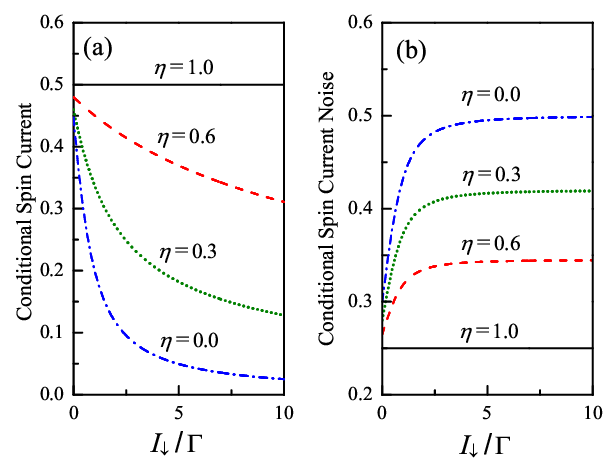}
 \caption{\label{fig4}
 (a) Conditional spin-$\upa$ current and (b) its noise as functions of
 given spin-$\dwa$ currents for different values of $\eta$.
 The tunnel-couplings are symmetric, i.e. $\GamL=\GamR=1.0$. The
 currents are measured in unit of $\Gam=\GamL+\GamR$.}
 \end{center}
 \end{figure}

 \section{\label{thsec5}Results and Discussion}

 First let us focus on the regime (i), as schematically shown
 in \Fig{Fig1}(a).
 The involving states are $|0\ra$-empty dot,
 $|\!\!\upa/\dwa\ra$-occupation by a spin-$\upa\!/\!\dwa$
 electron, and $|\rmd\ra$-doubly occupied.
 The ($\chi_\upa,\chi_\dwa$)-resolved quantum master
 equation reads $\frac{\partial}{\partial t}\rho(\chi_\upa,\chi_\dwa)
 =M_1(\chi_\upa,\chi_\dwa)\rho(\chi_\upa,\chi_\dwa)$, where
 $M_1$ is a 4$\times$4 matrix
 \bea
 M_1=\left(\begin{array}{cccc}
   -2\GamL & e^{\rmi\chi_\upa}\GamR & e^{\rmi\chi_\dwa}\GamR & 0 \\
   \GamL & -\GamL-\GamR & 0 & e^{\rmi\chi_\dwa}\GamR \\
   \GamL & 0 & -\GamL-\GamR & e^{\rmi\chi_\upa}\GamR \\
   0 & \GamL & \GamL  & -2 \GamR
   \end{array}\right).
 \eea
 The minimal eigenvalue of $M_1$ is then determined,
 which leads to the joint
 distribution $P_1(I_\upa,I_\dwa,t_{\rm c})$ of
 spin $\upa$ and $\dwa$ currents
 \be
 \frac{\log P_1}{t_{\rm c}}\!=\!-(\GamL+\GamR)+\!\sum_{\sgm=\upa,\dwa}\!
 \left\{\frac{\Omega_\sgm}{2}
 \!-\!I_\sgm\log\left(\frac{\Omega_\sgm I_\sgm}{\GamL\GamR}\right)\right\},
 \ee
 with $\Omega_\sgm=2I_\sgm+\sqrt{4I_\sgm^2+(\GamL-\GamR)^2}$.
 Apparently, the joint probability
 factorizes $P_1(I_\upa,I_\dwa)=P(I_\upa)P(I_\dwa)$,
 which implies that spin-$\upa$ and spin-$\dwa$ currents are uncorrelated.
 The resultant mixed generating function for conditional spin counting
 statistics then reads
 \be
 {\cal F}_1(\chi_\upa|I_\dwa)\!=\!-\frac{t_\rmc}{2}\{\GamL+\GamR
 -\sqrt{(\GamL\!-\!\GamR)^2\!+\!4\GamL\GamR e^{-\rmi \chi_\upa}}\},
 \ee
 analogous to the charge counting statistics in the regime
 (i) in \Eq{CGFI}, except for an overall factor of $\frac{1}{2}$.
 The spin transport through the QD
 independently, thus each spin component contributes
 50\% of the total current.
 The cumulants of conditional spin counting statistics
 are thus the same as those of unconditional ones. For
 instance, the first and second mixed cumulants are
 given, respectively, by
 \begin{gather}\label{cumuModI}
 \la I_\upa\ra_{I_\dwa}=\frac{\GamL\GamR}{\GamL+\GamR},
 \\
 \frac{\la I^2_\upa\ra_{I_\dwa}}{2e\la I_\upa\ra_{I_\dwa}}
 =1-\frac{2\GamL\GamR}{(\GamL+\GamR)^2}.
 \end{gather}
 Thus, in this case the system can be mapped onto that
 transport through a single level without Coulomb interaction.

 \begin{figure}
 \begin{center}
 \includegraphics*[scale=0.8]{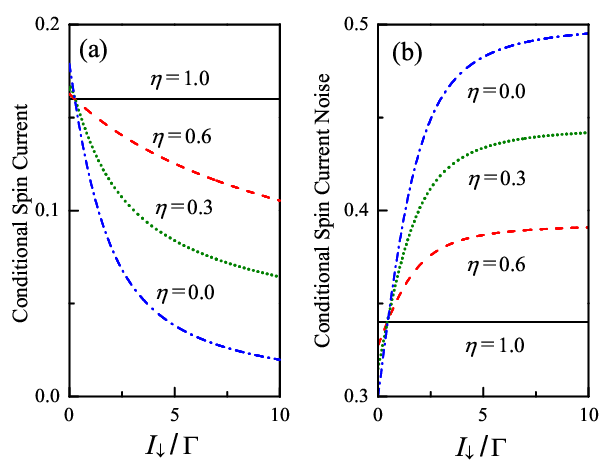}
 \caption{\label{fig5}
 Conditional spin-$\upa$ current and its noise versus $I_\dwa$
 for asymmetric tunnel-couplings $\GamR=4\GamL$.
 The other parameters are the same as those in \Fig{fig4}.}
 \end{center}
 \end{figure}

 Let us now consider the conditional spin counting statistics
 in the regime (ii) as shown in \Fig{Fig1}(b), where at most one
 electron can be occupied and thus up and down spins are maximally
 correlated.
 The involving states are reduced to $|0\ra$-empty dot,
 $|\!\upa$/$\dwa\ra$-occupied by a spin $\upa$/$\dwa$
 electron.
 The spin-resolved quantum master
 equation reads $\frac{\partial}{\partial t}\rho(\chi_\upa,\chi_\dwa)
 =M_2(\chi_\upa,\chi_\dwa)\rho(\chi_\upa,\chi_\dwa)$,
 where $M_2$ is the 3$\times$3 matrix, given by
 \bea\label{MII}
 M_2=\left(\begin{array}{ccc}
   -2\GamL & e^{\rmi\chi_\upa}\GamR & e^{\rmi\chi_\dwa}\GamR \\
   \GamL & -\GamR & 0\\
   \GamL & 0 & -\GamR
   \end{array}\right).
 \eea
 The minimal eigenvalue can be readily evaluated, which
 results in the joint probability for the spin currents
 \bea
 \frac{\log P_2}{t_\rmc}=\frac{1}{2}(\Lambda-2\GamL-\GamR)
 -\sum_\sgm I_\sgm\log\left(\frac{\Lambda I_\sgm}{\GamL\GamR}\right),
 \eea
 where
 $\Lambda=2(I_\upa+I_\dwa)+\sqrt{4(I_\upa+I_\dwa)^2+(2\GamL-\GamR)^2}$.
 The term inside the logarithm shows unambiguously that the
 spin-$\upa$ and spin-$\dwa$ currents are correlated.
 Further integrating over the $\chi_\upa$ variable gives rise to
 the mixed generating function
 \be
 {\cal F}_2(\chi_\upa|I_\dwa)\!=\!\frac{t_\rmc}{2}
 \left\{G(\chi_\upa)\!-\!G(0)\!-\!2I_\dwa \log\!
 \left[\frac{2I_\dwa\!+\!G(\chi_\upa)}{2I_\dwa+G(0)}\right]\right\},
 \ee
 with $G({\chi_\upa})=\sqrt{4I_\dwa^2
 + 4 e^{\chi_\upa}\GamL\GamR+(2\GamL-\GamR)^2}$.
 The conditional spin counting statistics can be
 calculated via taking derivatives with respective
 to $\chi_\upa$. For instance, the first mixed
 cumulant yields the conditional spin current
 \bsube
 \be\label{CIP}
 \la I_\upa\ra_{I_\dwa}=\frac{\GamL\GamR}
 {2I_\dwa+\Gup},
 \ee
 and the second mixed cumulant corresponds to
 the conditional shot noise of spin $\upa$ current
 \be\label{CSN}
 \frac{\la I^2_\upa\ra_{I_\dwa}}{e\la I_\upa\ra_{I_\dwa}}=
 1-\frac{2\GamL\GamR }{[2I_\dwa + \Gup]\Gup},
 \ee
 \esube
 where $G(0)=\sqrt{4I_\dwa^2+4\GamL^2+\GamR^2}$.
 Both conditional current and noise show radical changes in
 comparison with the uncorrelated ones [see \Eq{cumuModI} for the
 regime (i)].

 It is also instructive to compare the conditional spin
 counting statistics with the unconditional one.
 By utilizing \Eq{MII}, the first two cumulants of the
 unconditional spin counting statistics are given, respectively,
 by
 \begin{gather}\label{UNSCS}
 \la I_\upa\ra_{\rm un}=\frac{\GamL\GamR}{2\GamL+\GamR},
 \\
 \frac{\la I^2_\upa\ra_{\rm un}}{e\la I_\upa\ra_{\rm un}}
 =1-\frac{2\GamL\GamR}{(2\GamL+\GamR)^2}.
 \end{gather}
 Except for an effective doubling of ``$\GamL$'', these
 two cumulants are found to be qualitatively
 analogous to the results in regime (i).
 We will shown, on the contrary, that the conditional spin
 counting statistics is much more sensitive to the Coulomb
 repulsion.
 In case of spin polarized transport, it will be revealed
 that the conditional spin counting statistics can be served
 as transparent tool to investigate spin-resolved bunching
 behavior in mesoscopic transport.

 \begin{figure}
 \begin{center}
 \includegraphics*[scale=0.75]{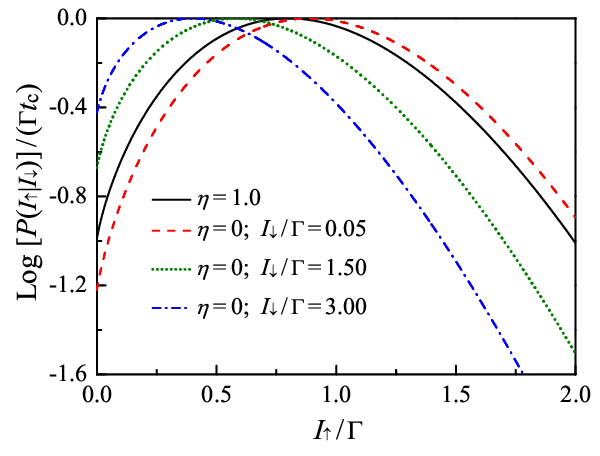}
 \caption{\label{fig6}
 Conditional spin current statistics versus $I_\upa$ for
 uncorrelated ($\eta=1.0$) and maximally correlated
 ($\eta=0.0$) transport.
 Other parameters are the same as those in \Fig{fig5}.}
 \end{center}
 \end{figure}

 We are now in a position to investigate the conditional
 spin counting statistics between the regimes (i) and (ii),
 as shown in \Fig{Fig1}(c).
 It thus offers an interpolation between the uncorrelated
 and the maximally correlated regimes.
 In experiments, it can be realized  via tuning appropriately
 the gate and bias voltages, shifting thus the excitation energies
 ($\epl_\sgm$ and $\epl_\sgm+U$) relative to the Fermi levels
 of the left and right electrodes in such a way that double
 occupation on the QD is partially allowed \cite{Ono021313}.
 The corresponding rate matrix is then given by
 \begin{widetext}
 \bea
 \hspace{-0.6cm}
 M_3=\left(\begin{array}{cccc}
   -2\GamL & e^{\rmi\chi_\upa}\GamR & e^{\rmi\chi_\dwa}\GamR & 0 \\
   \GamL & -\eta\GamL-\GamR & 0 & (1-\eta)\GamL+e^{\rmi\chi_\dwa}\GamR \\
   \GamL & 0 & -\eta\GamL-\GamR & (1-\eta)\GamL+e^{\rmi\chi_\upa}\GamR \\
   0 & \eta\GamL & \eta\GamL  & -2(1-\eta)\GamL-2\GamR
   \end{array}\right),\nonumber
 \eea
 \end{widetext}
 where $\eta=\{1+e^{\beta(\epl_\sgm+U-\mu_\rmL)}\}^{-1}$
 represents the Fermi function.
 If the excitation energy $\epl_\sgm+U$ is well below
 $\mu_\rmL$, $\eta$ approaches 1 and $M_3$ reduces to $M_1$,
 corresponding to uncorrelated transport.
 In the opposite limit of $\epl_\sgm+U\gg\mu_\rmL$, $\eta$
 goes to zero and $M_3\rightarrow M_2$,
 leading thus to the regime of maximal correlation.
 The rate matrix $M_3$ thus allows us to investigate an interpolation
 between the two extreme cases.

 The numerical results of conditional spin current and its shot
 noise are displayed in \Fig{fig4}(a) and (b), respectively,
 for symmetric tunnel-couplings ($\GamL=\GamR$).
 The conditioned spin-$\upa$ current decreases monotonically
 with the spin-$\dwa$ current, and tends to 0 as
 $I_\dwa\rightarrow\infty$.
 The conditional shot noise of spin-$\upa$ current, however,
 increases with $I_\dwa$.
 For a given $I_\dwa$, the spin-$\upa$ current grows with
 rising $\eta$, approaching the maximum at $\eta=1.0$
 corresponding to uncorrelated transport.
 Yet, the conditional noise decreases as $\eta$ rises, and
 reaches the minimum at $\eta=1.0$.
 Unambiguously, in the regime $I_\dwa/\Gam\gg1$,
 both conditional current and noise are very sensitive to $\eta$,
 thus showing them as sensitive tools to probe the Coulomb correlation.

 \Fig{fig5} shows the first two conditional cumulants
 for asymmetric tunnel-couplings ($\GamR=4\GamL$).
 The results are qualitatively similar to those in \Fig{fig4},
 except for the regime of small $I_\dwa$ ($I_\dwa/\Gam\lesssim1$),
 where finite Coulomb correlation leads to a spin-$\upa$ current
 exceeding the uncorrelated one ($\eta=1.0$).
 This can be explained in terms of the conditional spin current
 statistics $P(I_\upa|I_\dwa)$, as shown in \Fig{fig6}.
 Consider the situation of maximal correlation ($\eta=0$).
 For a given small spin-$\dwa$ current (see the dashed curve for
 $I_\dwa/\Gam=0.05$), the average of the conditional probability
 is slightly larger than that of uncorrelated transport
 ($\eta=1.0$), resulting thus in the unique behavior in the regime
 of small $I_\dwa$.
 With increasing $I_\dwa$, the average of the distribution
 decreases [see, for instance, the dotted curve for $I_\dwa/\Gam=1.5$].
 Eventually, it results in suppressed conditional current in
 comparison with the uncorrelated transport, as displayed in \Fig{fig5}.
 Interestingly, it is found that in the limit of maximal correlation
 ($\eta=0$), the conditional noise reaches the maximum $\frac{1}{2}$ as
 $I_\dwa\rightarrow\infty$, regardless of the
 ratio between the two tunneling rates $\GamL/\GamR$ [cf. \Eq{CSN}].
 It actually reflects the rare tunneling events of up spins
 through the QD system given an extremely large
 spin-$\dwa$ current.

 So far, our analysis has been focused only on the situation where
 the transport is spin unpolarized. Now we account for finite
 polarization due to ferromagnetic electrodes, which give rise to
 polarization-dependent tunneling rates.
 Consider, for instance, the parallel magnetic configuration when
 the majority of electrons in both electrodes point in the same direction.
 The rates of tunneling through the left and right electrodes
 are
 \be
 \Gam_{{\rm L}\upa/{\rm L}\dwa}=(1\pm p)\GamL
 \quad
 {\rm and}
 \quad
 \Gam_{{\rm R}\upa/{\rm R}\dwa}=(1\pm p)\GamR,
 \ee
 respectively, where we have assumed the same polarization $p$ $(0\leq p\leq1)$ in
 the left and right electrodes.
 In the limit of maximum correlation ($\eta=0$), the corresponding
 rate matrix is reduced to
 \bea\label{MIV}
 M_4=\left(\begin{array}{ccc}
   -2\GamL & e^{\rmi\chi_\upa}(1+p)\GamR & e^{\rmi\chi_\dwa}(1-p)\GamR \\
   (1+p)\GamL & -(1+p)\GamR & 0\\
   (1-p)\GamL & 0 & -(1-p)\GamR
   \end{array}\right).
 \eea
 The unconditional spin $\upa$ current and its noise are given
 by
 \bsube
 \bea
 \la I_\upa\ra_{\rm un}&\!\!\!=\!\!\!&(1+p)\frac{\GamL\GamR}{2\GamL+\GamR},
 \\
 \frac{\la I^2_\upa\ra_{\rm un}}{e\la I_\upa\ra_{\rm un}}
 &\!\!\!=\!\!\!&1-\frac{2\GamL\GamR}{(2\GamL+\GamR)^2}+
 \frac{4p\GamL^2}{(1-p)(2\GamL+\GamR)^2}.
 \eea
 \esube
 In the limit $p\rightarrow0$, one recovers the results in \Eq{UNSCS}.
 Both current and noise increase with the polarization $p$.
 In particular, the unconditional spin current noise may
 increase dramatically, and even diverge as $p\rightarrow1$.

 \begin{figure}
 \begin{center}
 \includegraphics*[scale=0.9]{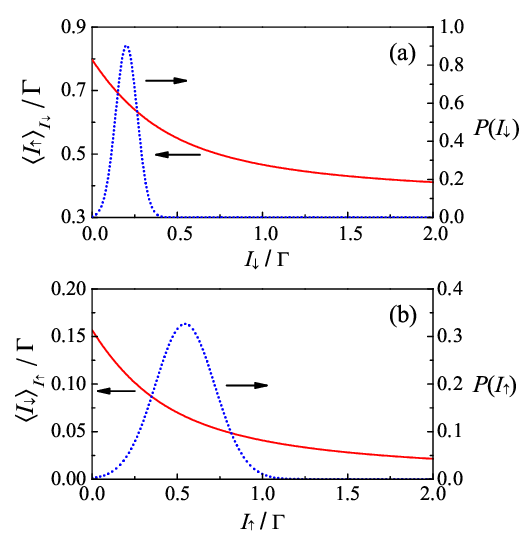}
 \caption{\label{fig7}
 (a) Conditional spin $\upa$ current, (b) spin $\dwa$ current, and
 their related probabilities [$P(I_\dwa)$ and $P(I_\upa)$]
 for polarization $p$=0.5 and symmetric tunnel-couplings $\GamL=\GamR=\Gam/2$.}
 \end{center}
 \end{figure}

 To further elucidate the underlying mechanism, we employ the
 conditional spin counting statistics.
 It is inferred from the Bayes theorem and \Eq{CSCS} that
 \be
 \la I_\upa\ra_{\rm un}=\int_0^\infty \rmd I_\dwa
 P(I_\dwa) \la I_\upa\ra_{I_\dwa},
 \ee
 where $P(I_\dwa)$ is the probability distribution of the
 spin $\dwa$ current that can be obtained from \Eq{MIV}.
 It allows us to analyze contribution to the unconditional
 spin current under various circumstances.
 \Fig{fig7} shows the numerical results of the conditional
 spin current and related probability for
 polarization $p=0.5$ and symmetric tunnel-couplings.
 It is illustrated unambiguously in \Fig{fig7}(a) that the
 major contribution to the unconditional spin $\upa$ current
 ($\la I_\upa\ra_{\rm un}$) comes from a large
 conditional spin $\upa$ current ($\la I_\upa\ra_{I_\dwa}\sim0.6\Gam$)
 at small spin $\dwa$ current in the narrow window
 defined by the probability $P(I_\dwa)$, i.e. $I_\dwa\sim(0.2\pm0.1)\Gam$.
 It reveals the bunching of up spins due to a dynamical
 blockade of a small spin $\dwa$ current, which eventually results in
 the prominent super-Poissonian noise as $p\rightarrow1$.
 On the contrary, for $\la I_\dwa\ra_{\rm un}$ the main contribution
 is from the low unconditional spin $\dwa$ current within a wide range
 of spin $\upa$ current  $I_\dwa\sim(0.5\pm0.3)\Gam$,
 as shown in \Fig{fig7}(b). The noise of spin $\dwa$ current is thus
 suppressed below the Poisson value, as we have checked.
 Despite this simple QD system, our analysis demonstrated
 that the conditional spin counting statistics is
 a sensitive and useful tool to investigate spin-resolved
 bunching behavior and its connection to the noise characteristics.

 Finally, let us propose a scheme to measure the conditional spin
 counting statistics.
 Instead of using the normal electrodes, we consider a
 four-terminal device: A single QD tunnel-coupled to
 four fully spin polarized ferromagnetic
 electrodes \cite{Sau04106601,San03214501}.
 The two left (right) electrodes are kept at the same chemical
 potential $\mu_{\rm L}$ ($\mu_{\rm R}$) but with opposite spin
 polarizations.
 The junction parameters are set the same for the two left
 electrodes, and likewise for the two on the right, such that
 the net current transport through the device is not spin
 polarized and the proposed four-terminal setup can be mapped onto the model
 we have analyzed.
 The great advantage of this scheme is that it is possible
 to measure separately the up or down spin current in each of
 the four electrodes, even though the net current is not spin
 polarized.
 Thus, it eventually enables us to measure the conditional spin
 counting statistics.
 We expect these predictions would be tested in quantum
 transport experiments in the near future.

 \section{\label{thsec6}Conclusion}

 In summary, we have proposed the theory of conditional spin
 counting statistics: The statistical fluctuations of one
 spin current component, given the observation of another spin
 current component.
 In the context of transport through a single quantum dot, we
 demonstrated that the presence of a Coulomb correlation leads
 to conditional spin counting statistics that exhibits a
 substantial change in comparison to the uncorrelated one.
 It thus can be served as an effective way to
 characterize the Coulomb correlation in mesoscopic transport
 systems.
 We further show that in the spin polarized transport, the
 conditional spin counting statistics offers a transparent
 and sensitive tool to understand the spin-resolved
 bunching behavior and its connection to the noise
 characteristics in mesoscopic transport.\cite{Ema12165417}.

 \begin{acknowledgements}
 This work was supported by the
 National Natural Science Foundation of China
 (grant Nos. 11204272, 11147114, and 11004124)
 and the Zhejiang Provincial Natural Science Foundation
 (grant Nos. Y6110467 and LY12A04008).
 \end{acknowledgements}


\end{document}